\documentclass[12pt,epsfig]{article}
\usepackage{epsfig}
\usepackage{amssymb}
\usepackage{amsmath}
\newcommand{\singlespace}{
     \renewcommand{\baselinestretch}{1}\large\normalsize}
\newcommand{\doublespace}{
     \renewcommand{\baselinestretch}{1.6}\large\normalsize}

\newcommand{\beq}{\begin{equation}}
\newcommand{\eeq}{\end{equation}}
\newcommand{\bea}{\begin{eqnarray}}
\newcommand{\eea}{\end{eqnarray}}

\newcommand{\ave}[1]{\langle {#1} \rangle}

\newcommand{\pslash}{p\!\!\!/}

\newcommand{\pb}{\bar\psi}
\newcommand{\qq}{\ave{\pb\psi}}
\newcommand{\eq}[1]{Eq.~(\ref{#1})}
\newcommand{\eqs}[1]{Eqs.~(\ref{#1})}

\def\roughly#1{\mathrel{\raise.3ex\hbox{$#1$\kern-.75em%
\lower1ex\hbox{$\sim$}}}}

\def\={\;=\;}
\def\+{\;+\;}

\begin{document}
%
\begin{flushright}
May 2001
\end{flushright}
\vspace{1.0cm}
\begin{center}
\doublespace
\begin{large}
{\bf }Self-consistent parametrization of the two-flavor isotropic
color-superconducting ground state\\
\end{large}
\vskip 1.0in
M. Buballa$^a$, J. Ho\v sek$^b$, and M. Oertel$^{c}$\\
{\small{\it $^a$ Institut f\"ur Kernphysik, TU Darmstadt,
                 Schlossgartenstr. 9, 64289 Darmstadt, Germany\\
            $^b$ Dept. Theoretical Physics, Nuclear Physics Institute,
                 25068 \v Re\v z (Prague), Czech Republic\\
            $^c$ IPN-Lyon, 43 Bd du 11 Novembre 1918,
                 69622 Villeurbanne C\'edex, France}}\\
\end{center}
\vspace{1cm}

\begin{abstract}
Lack of Lorentz invariance of QCD at finite quark chemical
potential in general implies the need of Lorentz non-invariant
condensates for the self-consistent description of the
color-superconducting ground state. Moreover, the spontaneous
breakdown of color $SU(3)$ in this state naturally leads to the
existence of $SU(3)$ non-invariant non-superconducting expectation
values. We illustrate these observations by analyzing the
properties of an effective 2-flavor Nambu--Jona-Lasinio type
Lagrangian and discuss the possibility of color-superconducting
states with effectively gapless fermionic excitations. It turns
out that the effect of condensates so far neglected can yield new
interesting phenomena.
\end{abstract}

\newpage

\singlespace

\section{Introduction}
According to current wisdom the deconfined 2-flavor QCD matter at moderate
baryon densities and low temperatures behaves as a color superconductor
\cite{RaWi00}.
Standard characteristics of its BCS-type ground state is the ground-state
expectation value \cite{AWR98,RSSV98}
\begin{equation}
     \delta_1 \= \ave{\psi^T \,C \gamma_5 \,\tau_2 \,\lambda_2 \,\psi} \;,
\label{delta1}
\end{equation}
which corresponds to a scalar diquark condensate in a color
anti-triplet state. Here the superscript $T$ denotes a transposition and
$C$ the matrix of charge conjugation.
$\tau_2$ is a Pauli matrix acting in flavor space,
$\lambda_2$ a Gell-Mann matrix acting in color space.
We have taken the freedom to rotate the unpaired quark color
into the 3-direction, which can always be done.

Because of the empirical fact that the (approximate) chiral SU(2)
symmetry of the QCD Lagrangian
 is not respected by the QCD vacuum, it is
natural to ask whether the quark (-antiquark) condensate
\begin{equation}
     \phi \=\ave{\pb \,\psi} \;,
\label{phi}
\end{equation}
persists also in the ground state of QCD matter at finite baryon
density. For model Lagrangians with exact chiral symmetry one
usually finds a first-order phase transition from a low-density
phase (or vacuum) with $\phi \neq 0$ and $\delta_1 = 0$ to a
high-density phase with $\delta_1 \neq 0$ and $\phi = 0$. This is
different if there is a small quark mass $m$ which explicitly
breaks chiral symmetry. In this case $\phi$ cannot exactly vanish
above the phase transition and coexists with the diquark
condensate. In fact, just above the phase transition the gaps
related to the two condensates can be of similar magnitude
\cite{BeRa99}.

At finite density the existence of Lorentz non-invariant expectation values
becomes possible. The most obvious example is of course the density itself,
\beq
     \rho \= \ave{\pb \,\gamma^0\,\psi} \;,
\label{rho}
\end{equation}
which transforms like the time component of a 4-vector.
Consequently, if the Pauli principle and the form of the interaction
permit, the ground state of a 2-flavor color superconductor is
characterized by more condensates than merely the Lorentz-invariant ones,
(\ref{delta1}) and (\ref{phi}). This simple observation was made
already in the pioneering work of Bailin and Love \cite{BaLo84}.
For instance, there could be another diquark condensate of the
form \cite{BaLo84,LaRh99,ABR99}
\beq
     \delta_2 \= \ave{\psi^T\,C\gamma^0\gamma_5\,\tau_2 \,\lambda_2 \,\psi}
     \;,
\label{delta2}
\end{equation}
which also transforms like the time component of a 4-vector.
The ground state does not even have to be isotropic, but rotational
invariance could be spontaneously broken, like in a ferromagnet.
Even the breakdown of translational invariance due to the formation
of crystalline phases is a possible scenario \cite{RSZ01,ABoR00}.

Up to now, we have discussed only space-time symmetries. We should
recall, however, that in the presence of the diquark condensates
(\ref{delta1}) or (\ref{delta2}) also color $SU(3)$ is broken,
since only the first two colors (``red'' and ``green'')
participate in the condensate, while the third one (``blue'') does
not. Therefore, there is no reason to assume that all other
condensates are color-$SU(3)$ invariant in this state. For
instance, we should expect that the contributions of red and blue
quarks, $\phi_r$ and $\phi_b$,  to the quark condensate $\phi$
could be different, thus giving rise to a non-vanishing
expectation value \beq
     \phi_8 \= \ave{\pb \,\lambda_8\,\psi}
            \= \frac{2}{\sqrt{3}}\,(\phi_r - \phi_b) \;.
\label{phi8}
\end{equation}
(Note that $\delta_1$ and $\delta_2$ leave a color $SU(2)$
subgroup invariant and therefore all green quantities are
identical to the red ones). Similarly, the densities of red and
blue quarks will in general not be the same, i.e., in addition to
the total number density $\rho = 2\rho_r + \rho_b$ there could be
a non-vanishing expectation value \beq
     \rho_8 \=\ave{\pb \,\gamma^0\,\lambda_8\,\psi}
            \= \frac{2}{\sqrt{3}}\,(\rho_r - \rho_b) \;.
\label{rho8}
\end{equation}
Since these color-symmetry breaking expectation values,
induced by the presence of color-symmetry breaking diquark condensates,
could in turn influence the properties of the diquark condensates,
in principle, all condensates should be studied in a self-consistent way.

Detailed understanding of the color-superconducting state which in
Nature might exist in the interiors of neutron stars \cite{We99}
requires, however, a detailed knowledge of the effective
quark-quark interactions close to their Fermi surface \cite{Po92}.
Due to the lack of experimental data or information from the
lattice such a knowledge is missing at present. It is nevertheless
possible to analyze in detail the effective low-energy quantum
field theory of the deconfined low (or zero) temperature quark
matter within models which respect all relevant symmetries of the
corresponding QCD Lagrangian. Lack of Lorentz invariance of QCD at
finite $\mu$ implies that ${\cal L}_{eff}$ itself should respect
only an $O(3)$ rotational symmetry. Since there are no thermal
gluons at $T=0$ it is justified and customary to analyze the
ground-state properties of deconfined quark matter by virtue of
${\cal L}_{eff}$ having only a global color $SU(3)$ symmetry.
Response of gluons to the quark condensates carrying color
(Meissner effect) is then studied \cite{Ri00,CaDi00}
perturbatively, i.e., as if the color gauge symmetry is
spontaneously broken (for more detailed discussion see
\cite{RaWi00}). In this article we illustrate the need of Lorentz-
and color-$SU(3)$ non-invariant condensates for the
self-consistent description of the color-superconducting ground
state. For simplicity we leave the rotational symmetry
unbroken, i.e., we restrict ourselves to the
isotropic case.

The remainder of this article is organized as follows. In
Sec.~\ref{formalism} we will derive the thermodynamic potential
and a coupled set of gap equations for a general NJL-type model,
taking into account a possible condensation in the channels
(\ref{delta1}) to (\ref{rho8}). The resulting dispersion laws are
discussed in more detail in Sec.~\ref{disp}. Here special emphasis
is put on possible gapless color superconductors. In
Sec.~\ref{results} we discuss our numerical results. Conclusions
are drawn in Sec.~\ref{conclusions}.

\section{General Formalism}
\label{formalism}

In order to illustrate the generic properties of the superconducting
ground state let us -as an example- consider a model
defined by the Lagrangian density
\beq
    {\cal L}_{eff} \= \pb (i \partial\hspace{-2.3mm}/ - m) \psi
                      \+ {\cal L}_{q\bar q} \+ {\cal L}_{qq}
\label{Leff}
\end{equation}
describing a quark of mass $m$ which interacts with other quarks and
antiquarks via NJL-type 4-point interactions of the form
\beq
     {\cal L}_{q\bar q} \= g_s^{(0)}\,(\pb \psi)^2
                        \+ g_s^{(8)}\,(\pb \lambda_a\psi)^2
                        \+ g_v^{(0)}\,(\pb \gamma^0\psi)^2
                        \+ g_v^{(8)}\,(\pb \gamma^0 \lambda_a\psi)^2
                        \+ ...
\label{Lqqbar}
\end{equation}
and
\beq
     {\cal L}_{qq} \= h_1\,(\pb \,i\gamma_5 \tau_2 \lambda_A \,\psi^c)
                           (\pb^c \,i\gamma_5 \tau_2 \lambda_A \, \psi)
     \+  h_2\,(\pb \,\gamma^0\gamma_5 \tau_2 \lambda_A \,\psi^c)
              (\pb^c \,\gamma^0\gamma_5 \tau_2 \lambda_A \, \psi)
     \+ ...
\label{Lqq}
\end{equation}
Here the dots indicate possible other channels, not related to
the expectation values (\ref{delta1}) to (\ref{rho8}).
In \eq{Lqq} we used the notation $\psi^c = C \pb^T$, while $\lambda_A$
denotes the antisymmetric Gell-Mann matrices $\lambda_2$, $\lambda_5$, and
$\lambda_7$. All color indices are understood to be summed over.

In the following it is convenient to formally double the degrees of freedom
by defining
\beq
     q(x) \= \frac{1}{\sqrt 2}\,\left(\begin{array}{c} \!\!\psi(x)\!\! \\
     \!\!\psi^c(x)\!\! \end{array}\right)~.
\label{q}
\end{equation}
To obtain the mean-field thermodynamic potential at temperature $T$ and
chemical potential $\mu$ we linearize ${\cal L}_{eff}$ in the vicinity of
the expectation values \eqs{delta1} to (\ref{rho8}) and apply
Matsubara formalism.
The resulting thermodynamic potential per volume reads
\beq
    \Omega(T,\mu) \= -T \sum_n \int \frac{d^3p}{(2\pi)^3} \;
    \frac{1}{2}\,{\rm Tr}\; \ln \Big(\frac{1}{T}\,S^{-1}(i\omega_n, \vec p)
    \Big)
    \+ V \;,
\label{Omega}
\end{equation}
with $\omega_n$ being fermionic Matsubara frequencies.
$S^{-1}(p)$ is the inverse propagator of the q-fields at 4-momentum $p$.
It is given by
\beq
    S^{-1}(p) \= \left(\begin{array}{cc}
    \pslash - M_0 - M_8\lambda_8 + \mu_0\gamma^0 + \mu_8\gamma^0\lambda_8 &
    \Delta_1 \gamma_5\tau_2\lambda_2
    + \Delta_2 \gamma^0\gamma_5\tau_2\lambda_2 \\
    -\Delta_1^* \gamma_5\tau_2\lambda_2
    + \Delta_2^* \gamma^0\gamma_5\tau_2\lambda_2  &
    \pslash - M_0 - M_8\lambda_8 - \mu_0\gamma^0 - \mu_8\gamma^0\lambda_8
    \end{array}\right)\,.
\label{Sinv}
\end{equation}
Here we introduced the effective quark masses
\beq
    M_0 \= m - 2g_s^{(0)}\phi~,\qquad
    M_8 \= -2g_s^{(8)}\phi_8~,
\label{M08}
\end{equation}
the effective chemical potentials
\beq
    \mu_0 \= \mu + 2g_v^{(0)}\rho~,\qquad
    \mu_8 \;\= 2g_v^{(8)}\rho_8~,
\label{mu08}
\end{equation}
and the diquark gaps
\beq
    \Delta_1 \= -2h_1 \delta_1~,\qquad
    \Delta_2 \= 2h_2 \delta_2~.
\label{Delta12}
\end{equation}
These quantities also enter into the last term of \eq{Omega},
which is defined as
\beq
    V \= \frac{(M_0 - m)^2}{4g_s^{(0)}} \+ \frac{M_8^2}{4g_s^{(8)}}
      \+ \frac{(\mu_0 - \mu)^2}{4g_v^{(0)}} \+ \frac{\mu_8^2}{4g_v^{(8)}}
      \+ \frac{|\Delta_1|^2}{4h_1} \+ \frac{|\Delta_2|^2}{4h_2} \;.
\label{pot}
\end{equation}
For later convenience, but also for the interpretation of the results,
it is useful to perform linear combinations to get red and blue quantities,
e.g. red and blue constituent quark masses
$M_r = M_0 + \frac{1}{\sqrt{3}}\,M_8$ and
$M_b = M_0 - \frac{2}{\sqrt{3}}\,M_8$.
We then find
\begin{alignat}{3}
  M_r \=& m &\;-\;& &\frac{2}{3}(6g_s^{(0)} &+ 2g_s^{(8)})\phi_r
             \;-\; \frac{2}{3}(3g_s^{(0)} - 2g_s^{(8)})\phi_b\,,
\nonumber \\
  M_b \=& m &\;-\;& &\frac{2}{3}(6g_s^{(0)} &- 4g_s^{(8)})\phi_r
             \;-\; \frac{2}{3}(3g_s^{(0)} + 4g_s^{(8)})\phi_b\,,
\nonumber \\
  \mu_r \=& \mu &\+& &\frac{2}{3}(6g_v^{(0)} &+ 2g_v^{(8)})\rho_r
                 \+ \frac{2}{3}(3g_v^{(0)} - 2g_v^{(8)})\rho_b\,,
\nonumber \\
  \mu_b \=& \mu &\+& &\frac{2}{3}(6g_v^{(0)} &- 4g_v^{(8)})\rho_r
                 \+ \frac{2}{3}(3g_v^{(0)} + 4g_v^{(8)})\rho_b\,.
\label{gapsr}
\end{alignat}

For two flavors and three colors the inverse propagator,
\eq{Sinv},  is a $48\times48$ matrix, and the trace in \eq{Omega}
has to be taken in this 48-dimensional space. After performing the
Matsubara sum we obtain:
\begin{alignat}{2}
    \Omega(T,\mu) \= -4 \int \frac{d^3p}{(2\pi)^3} \;\Big\{\quad
    2\hspace{5mm} \Big(&\frac{E_+ + E_-}{2} &
            &+\; T\,\ln(1 + e^{-E_+/T}) \+ T\,\ln(1 + e^{-E_-/T})
      \Big) \nonumber \\
           +\quad\Big(& \quad \epsilon_b &
            &+\; T\,\ln(1 + e^{-\epsilon_+/T})
              \+ T\,\ln(1 + e^{-\epsilon_-/T})
      \Big) \Big\} \nonumber \\
    \+ V\,, \hspace{3.5cm} &
\label{Omegaexp}
\end{alignat}
where physically irrelevant constant terms have been suppressed.
The dispersion laws which enter into this expression are given by
\beq
    \epsilon_\pm \= \epsilon_b \pm \mu_b
                 \= \sqrt{{\vec p}^2+ M_b^2} \pm \mu_b
\label{epspm}
\end{equation}
and
\beq
    E_\pm \= \sqrt{ {{\vec p}^2 + M_r^2 + \mu_r^2 + |\Delta_1|^2
                    + |\Delta_2|^2 \pm 2s}}\,,
\label{epm}
\end{equation}
with
\beq
    s \= \sqrt{(\mu_r^2 + |\Delta_2|^2){\vec p}^2 + t^2}\,,\qquad
    t \= M_r \mu_r - Re(\Delta_1\Delta_2^*) \,.
\end{equation}
So far, the thermodynamic potential depends on our choice of the
expectation values (\ref{delta1}) to (\ref{rho8}), which determine
the effective masses, effective chemical potentials and diquark
gaps as indicated above. On the other hand, in a thermodynamically
consistent treatment the condensates should follow from the
thermodynamic potential by taking the appropriate derivatives. The
self-consistent solutions are given by the stationary points of
the potential, \beq
      \frac{\delta\Omega}{\delta M_0} \= \frac{\delta\Omega}{\delta M_8}
   \= \frac{\delta\Omega}{\delta\mu_0} \= \frac{\delta\Omega}{\delta\mu_8}
   \= \frac{\delta\Omega}{\delta\Delta_1}
   \= \frac{\delta\Omega}{\delta\Delta_2} \= 0 \,.
\label{stat}
\end{equation}
If there is more than one stationary point, the stable solution is the one
which corresponds to the lowest value of $\Omega(T,\mu)$.

\eqs{Omegaexp} and (\ref{stat}) lead to the following expressions for the
various expectation values:
\begin{alignat}{2}
    &\phi_r \=  & -4 \int \frac{d^3p}{(2\pi)^3}\; \frac{1}{2s} \;\Big\{
    \quad (1-2n(E_+))\; &\frac{1}{E_+}\,
    [ M_r s + \mu_r t ]\quad
\nonumber \\
    & & +\;(1-2n(E_-))\; &\frac{1}{E_-}\,
    [ M_r s - \mu_r t ]\; \Big\}\,,
\nonumber \\
    &\phi_b \= &-4 \int \frac{d^3p}{(2\pi)^3}\; \frac{M_b}{\epsilon_b} \;
    (1 - n(\epsilon_+) - n(&\epsilon_-))\,,
\nonumber \\
    &\rho_r \= &4 \int \frac{d^3p}{(2\pi)^3}\; \frac{1}{2s} \;\Big\{
    \quad (1-2n(E_+))\; &\frac{1}{E_+}\,
    [ \mu_r(s+{\vec p}^2) + M_r t ]\quad
\nonumber \\
    & & +\;(1-2n(E_-))\; &\frac{1}{E_-}\,
    [ \mu_r(s-{\vec p}^2) - M_r t ]\;\Big\}\,,
\nonumber \\
    &\rho_b \= &4 \int \frac{d^3p}{(2\pi)^3}\; (n(\epsilon_-) - n(\epsilon_+))
    \,,
    \hspace{5mm}&
\nonumber \\
    &\delta_1 \= & -4 \int \frac{d^3p}{(2\pi)^3}\; \frac{1}{s} \;\Big\{
    \quad (1-2n(E_+))\; &\frac{1}{E_+}\,
    [ \Delta_1 s - \Delta_2 t ]\quad
\nonumber \\
    & & +\;(1-2n(E_-))\; & \frac{1}{E_-}\,
    [ \Delta_1 s + \Delta_2 t ]\; \Big\}\,,
\nonumber \\
    &\delta_2 \= & 4 \int \frac{d^3p}{(2\pi)^3}\; \frac{1}{s} \;\Big\{
    \quad (1-2n(E_+))\; &\frac{1}{E_+}\,
     [ \Delta_2 (s+{\vec p}^2) - \Delta_1 t ]
    \quad
\nonumber \\
    & & +\;(1-2n(E_-))\; &\frac{1}{E_-}\,
     [ \Delta_2 (s-{\vec p}^2) + \Delta_1 t ]
    \; \Big\}\,,
\label{gap}
\end{alignat}
where $n(E) = 1/(e^{E/T} + 1)$ is a Fermi function. Together with
\eqs{Delta12} and (\ref{gapsr}) these equations form a set of six
coupled gap equations for $M_r$, $M_b$, $\mu_r$, $\mu_b$,
$\Delta_1$ and $\Delta_2$. The expressions for the blue
expectation values $\phi_b$ and $\rho_b$ formally look like the
corresponding formulae for free particles. However, they depend on
the effective quantities $M_b$ and $\mu_b$, which are also related
to red quantities via \eq{gapsr}. Despite of these
interdependencies, the masses of red and blue quarks, and also
their densities will in general be different, as anticipated
above.

Another interesting observation is that in general $\delta_1$ and $\delta_2$
or, equivalently, $\Delta_1$ and $\Delta_2$ cannot vanish separately.
This means that the familiar scalar diquark condensate $\delta_1$ will in
general be accompanied by an induced non-vanishing expectation value
$\delta_2$. \footnote{This point has already been discussed in
Ref.~\cite{ABR99}. The authors, however, argued that because of the small
value of $\delta_2$ as compared with $\delta_1$, this condensate could for
simplicity be neglected although strictly speaking in this case the system of gap
equations cannot be closed.}
On the other hand, there is always a solution with $\Delta_1=\Delta_2 = 0$.
In this case the expressions for $\phi_r$ and $\rho_r$ get the same
structure as the analogous expressions for the blue quarks.
If $\Delta_1$ and $\Delta_2$ do not vanish, they are determined by the
gap equation only up to a common phase. In most cases this phase can be
chosen such that $\Delta_1$ and $\Delta_2$ are real.

\section{Dispersion laws and gapless color-superconducting states}
\label{disp}

In physical terms the derived mean-field thermodynamic potential,
\eq{Omegaexp}, describes the thermodynamics of a mixture of
noninteracting fermionic excitations of two types: (A) quark
excitations with the dispersion law $\epsilon_{\pm}$ and (B)
Bogoliubov-Valatin (BV) quasiquark excitations with the dispersion
law $E_\pm$.\footnote{Strictly speaking, the quarks of component
(A) are also quasiparticles as long as $M_b$ is different from the
bare mass $m$.} Straightforward interpretation is to say that
these dispersion laws should be used for calculating (in
principle) measurable quantities, e.g. the specific heat of the
system. The parameters entering the dispersion laws, i.e.,
$M_{b}$, $M_{r}$, $\mu_{b}$, $\mu_{r}$, $|\Delta_1|$,
$|\Delta_2|$, and $\cos(\varphi_1 - \varphi_2)$, ($\Delta_{j} =
|\Delta_{j}| \exp{(i\varphi_{j})}, j=1,2$) are fixed as the
solutions of the coupled nonlinear integral equations (\ref{gap})
in terms of $m$, $\mu$, and the dimensional coupling constant(s)
of a given four-fermion interaction. Due to the fact that the
integrals in \eq{gap} have to be regularized there is one more
dimensionful parameter upon which the above quantities depend.
This can be a cutoff, a parameter in a form-factor \cite{AWR98},
the instanton size \cite{RSSV00}, or a thickness of a layer around
the Fermi surface in which the interaction is assumed to be
different from zero \cite{RSSV98}.

Component (A) of the system behaves like a normal relativistic Fermi
gas, characterized by a linear dependence of the low-temperature specific
heat on $T$. The actual behavior of the physically interesting BV component
of the system characterized by $E_{-}(\vec p)$ depends strongly upon the
details of the interaction.
For instance, the instanton mediated Lagrangians employed in
Refs.~\cite{AWR98} and \cite{RSSV98, RSSV00} have $h_2=0$ and therefore
$\Delta_2 = 0$.
In this case \eq{epm} reduces to the ``classic'' result
\begin{equation}
E_{-}(\vec p) = \sqrt{(\sqrt{\vec p^2 + M_{r}^2} - \mu_{r})^2 +
|\Delta_1|^2}\,.
\end{equation}
There are no nodes in this function, and this component exhibits
a superconducting behavior including the exponential low-$T$ specific heat
characteristic of ordinary superconductors.

In the general case the form of $E_-(\vec p)$ is given by \eq{epm}.
Clearly, without Lorentz invariance the dependence of the energy of a
particle-like excitation on momentum is restricted only by positivity.
It is not surprising, however, that $E_{-}(\vec p)$ can be parametrized as
\begin{equation}
E_{-}(\vec p) \equiv \sqrt{(\sqrt{\vec p^2 + M_{eff}^2} - \mu_{eff})^2 +
|\Delta_{eff}|^2}\,,
\end{equation}
provided we identify
\begin{equation}
\mu_{eff} \equiv \sqrt{\mu_{r}^2 + |\Delta_2|^2}\,,\quad
M_{eff}^2 \equiv \frac{(M_{r} \mu_{r} - Re \Delta_1\Delta_2^*)^2}
                      {\mu_{eff}^2}\,,
\label{mueff}
\end{equation}
and
\begin{equation}
M_{r}^2 + |\Delta_1|^2 \equiv M_{eff}^2 + |\Delta_{eff}|^2\,.
\label{Deltaeff}
\end{equation}
It is interesting to notice that $|\Delta_{eff}|$ can vanish,
i.e., $E_{-}(\vec p)$ can have nodes at $\vec p^2 = \mu_{eff}^2 -
M_{eff}^2$. This is the case if \beq
    M_r \Delta_2 = -\mu_r \Delta_1 \;.
\label{condition}
\end{equation}
If $\mu_r \neq 0$ this relation can be used to eliminate
$\Delta_1$ in \eq{mueff}:
\beq
    \mu_{eff}^2 = \mu_r^2 \, (1 + \frac{|\Delta_2|^2}{\mu_r^2})
    \,, \quad
    M_{eff}^2 = M_r^2 \, (1 + \frac{|\Delta_2|^2}{\mu_r^2}) \,,
\end{equation}
and hence
\beq
    \vec p_{node}^2 = (\mu_r^2 - M_r^2)\, (1 + \frac{|\Delta_2|^2}{\mu_r^2})
    \,.
\end{equation}
This means, $\mu_r^2$ must be greater or equal to $M_r^2$ and it
immediately follows from \eq{condition} that a gapless
color-superconducting  solution is only possible if $|\Delta_2|
\geq |\Delta_1|$.

In the vicinity of the node the BV quasiparticle takes the form of
a non-relativistic fermion,
\begin{equation}
E_{-}(\vec p) \approx \frac{\vec p^2}{2m^*} - \mu^* \,,
\end{equation}
where $m^*=\mu_{eff}$ and
$\mu^* = \frac{1}{2}(\mu_{eff} -\frac{M_{eff}^2}{\mu_{eff}})$.
Despite the superconducting condensates the specific heat of such
a system is linear in $T$. A similar phenomenon was found in
Ref.~\cite{ABeR00} in the color-flavor locked (CFL) phase, though its origin
is not the same as here.

In the above discussion it is tacitly assumed that there is an
interaction which yields solutions of the coupled gap equations
for which \eq{condition} holds.
A particularly simple way to fulfill this condition is to assume that
$M_r = \Delta_1 = 0$, which could be realized by taking
$m = g_i^{(k)} = h_1 = 0$ and only $h_2 \neq 0$.
If we regularize the divergent integrals by a sharp 3-momentum cut-off
$\Lambda$ and restrict ourselves to $T=0$ , the thermodynamic potential
of this schematic model is readily calculated:
\beq
    \Omega_{schem}(T=0,\mu;\Delta_2) \=
    -\frac{1}{6\pi^2} \Big( \;2(\mu^2 + |\Delta_2|^2)^2 + \mu^4 \; \Big)
    \+ \frac{|\Delta_2|^2}{4h_2} \,.
\end{equation}
Here we dropped an irrelevant constant
$-\frac{3}{2\pi^2}\Lambda^4$. Obviously this function is not
bounded from below, but only the self-consistent solutions of the
gap equation, i.e. $\delta\Omega_{schem} /\delta \Delta_2 = 0$,
are physically meaningful. There is always a trivial solution with
$\Delta_2 = 0$. For $0 < h_2 < \frac{3\pi^2}{8\mu^2}$ there are
also nontrivial solutions with $|\Delta_2|^2 = \frac{3\pi^2}{8h_2}
- \mu^2$. However, whenever these nontrivial solutions exist, they
correspond to maxima of $\Omega_{schem}$, while at same time the
trivial solution is a local minimum with a lower value of
$\Omega_{schem}$. This means, the nontrivial gapless solution is
unstable. Although we have shown this only for our very simple
schematic model, it might be a rather general feature. In fact, a
similar observation was made by the authors of Ref.~\cite{ABeR00}
for the CFL phase. We will come back to this point in the end of
the next section.

\section{Numerical results}
\label{results}

In the schematic example discussed in the end of the previous
section the problem was reduced to a single condensate. In this
section we want to present the results of a numerical study of the
full coupled set of gap equations derived in Sec.~\ref{formalism}.
As an example we consider an interaction with the structure of a
``heavy-gluon exchange'',
\beq
     {\cal L}_{hge} \= \pb (i \partial\hspace{-2.3mm}/ - m) \psi
              \;-\; g_E \,(\pb \gamma^0 \lambda_a \psi)^2
              \+ g_M \,(\pb \vec\gamma \lambda_a \psi)^2 \;,
\label{Lhge}
\end{equation}
although there is no reason why the effective interaction at
moderate densities should have this particular form. As discussed
in the introduction, the effective Lagrangian at finite densities
does not need to be Lorentz invariant. We underline this
possibility by explicitly allowing for different ``electric'' and
``magnetic'' coupling constants, $g_E$ and $g_M$.

The effective quark-antiquark interaction ${\cal L}_{q \bar q}$ and the
effective quark-quark interaction ${\cal L}_{qq}$ as given in \eqs{Lqqbar}
and (\ref{Lqq}) can be derived from ${\cal L}_{hge}$ by performing the
appropriate Fierz transformations.
The resulting coupling constants are
\begin{alignat}{3}
    g_s^{(0)} &= \frac{2}{9}\,(g_E+3g_M)~,\quad &
    g_s^{(8)} &= -\frac{1}{24}\,(g_E+3g_M)~,\quad&
    h_1 &= \frac{1}{6}\,\,(g_E+3g_M)~, \nonumber \\
    g_v^{(0)} &= \frac{2}{9}\,(g_E-3g_M)~,\quad &
    g_v^{(8)} &= -g_E-\frac{1}{24}\,(g_E-3g_M)~,\quad&
    h_2 &= \frac{1}{6}\,\,(g_E-3g_M)~. \nonumber \\
\label{geff}
\end{alignat}

\subsection{Lorentz invariant interaction}
\label{resultsli}

We begin our discussion with the case of a Lorentz-invariant interaction,
\beq
    g \;\equiv\; g_E \= g_M\,.
\label{lii}
\end{equation}
Although there is no reason for this assumption,
it allows for a better comparison with other calculations in the literature,
which often start from an interaction of this form. Of course, having
a Lorentz-invariant interaction does not mean that there are only
Lorentz-invariant condensates, since Lorentz-invariance is still broken
by the chemical potential.

If we insert \eq{lii} into \eq{geff} we find that for $g > 0$
the interaction is attractive in the scalar
quark-antiquark channel and in the scalar diquark channel and repulsive
in all other channels of interest.
Of course, non-vanishing expectation values in the repulsive channels
do not develop spontaneously, but only as a result of an external
source, like the chemical potential, or induced by non-vanishing
expectation values in attractive channels.
In fact, the solutions of the gap equations correspond to maxima of the
thermodynamic potential with respect to variations in the repulsive channels,
whereas they can be maxima or minima with respect to variations in the
attractive channels.

In our numerical calculations we restrict ourselves again to $T$~=~0
and take a sharp 3-momentum cut-off $\Lambda$ to regularize the integrals.
Although a quark model of the present type should be used only in the
deconfined phase, we follow the general custom and fix the parameters such
that they yield ``reasonable'' vacuum properties.
In the following we take $\Lambda$~=~600~MeV, $g\Lambda^2$~=~2.75,
and a bare quark mass $m$~=~5~MeV. With these parameter values
we obtain a vacuum constituent quark mass $M_r = M_b$~=~407.7~MeV.
This corresponds to a quark condensate $\phi = -2(245.7 {\rm MeV})^3$,
while $\phi_8$, $\rho$, $\rho_8$, $\delta_1$ and $\delta_2$ vanish in vacuum.

\begin{figure}
\epsfig{file=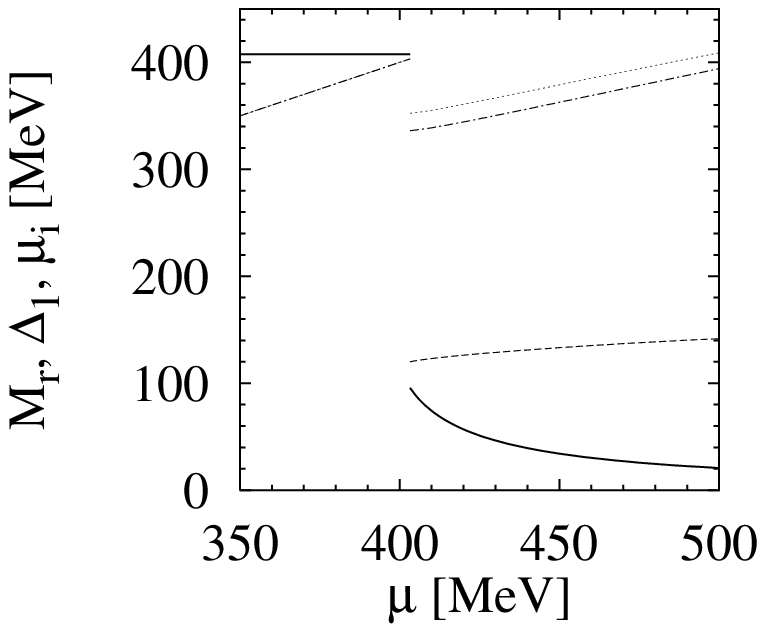,width=7.0cm}
\hfill
\epsfig{file=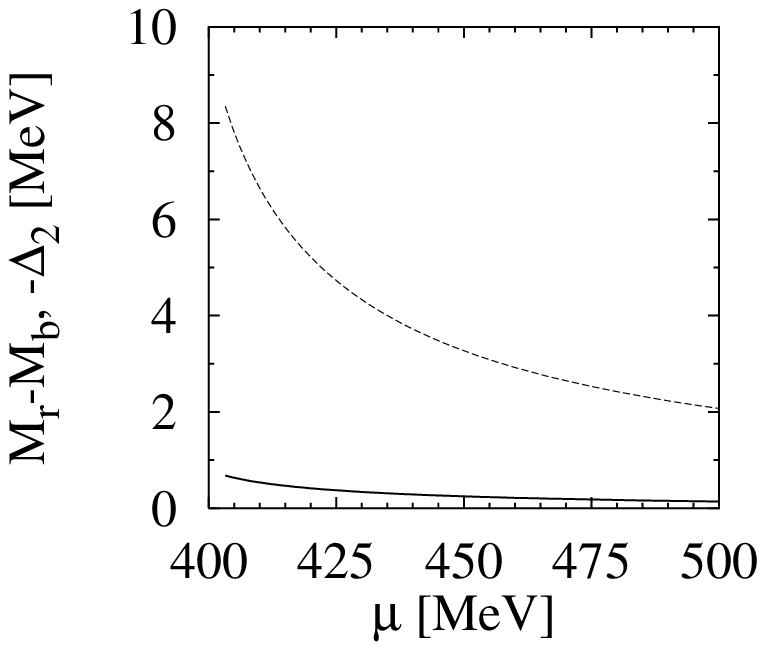,width=7.0cm}
\caption{\small Various quantities obtained with the Lorentz invariant
                interaction $g_E = g_M = 2.75/\Lambda^2$
                as functions of the quark chemical potential $\mu$.
                Left panel: $M_r$ (solid), $\Delta_1$ (dashed), $\mu_r$
                (dashed-dotted), $\mu_b$ (dotted).
                Right panel: $M_r-M_b$ (solid), $-\Delta_2$ (dashed).}
\label{gapsfig}
\end{figure}

When we increase the quark chemical potential nothing happens up
to a critical value $\mu_{crit}$~=~403.3~MeV. At this point a
first-order phase transition takes place and all expectation
values \eqs{delta1} to (\ref{rho8}) receive non-vanishing values.
This can be inferred from Fig.~\ref{gapsfig} where various
quantities are displayed as functions of $\mu$. In the left panel
we show the constituent quark mass $M_r$, the diquark gap
$\Delta_1$, and the effective chemical potentials $\mu_r$ and
$\mu_b$. In the right panel the mass difference $M_r-M_b$ and the
diquark gap $-\Delta_2$ are plotted. At $\mu=\mu_{crit}$ the
constituent quark masses drop by more than 300~MeV and are no
longer identical. The difference, however, is small,
$M_r$~=~95.7~MeV and $M_b$~=~95.0~MeV. With increasing chemical
potential, both, the masses and their difference, are further
decreased. In the diquark channel we find $\Delta_1$~=~120.0~MeV
at $\mu=\mu_{crit}$. Similar to what has been found in
Ref.~\cite{ABR99}, the second gap parameter has the opposite sign
and is more than one order of magnitude smaller,
$\Delta_2$~=~-8.4~MeV. Like the constituent masses, it decreases
with increasing $\mu$, whereas $\Delta_1$ is slightly growing in
the regime shown in Fig.~\ref{gapsfig}.

\begin{figure}
\begin{center}
\epsfig{file=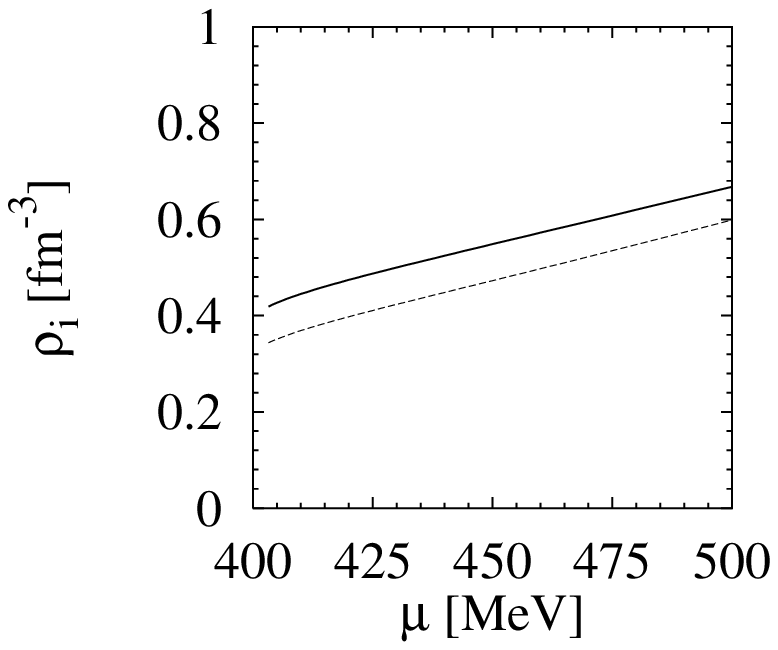,width=7.0cm}
\hfill
\epsfig{file=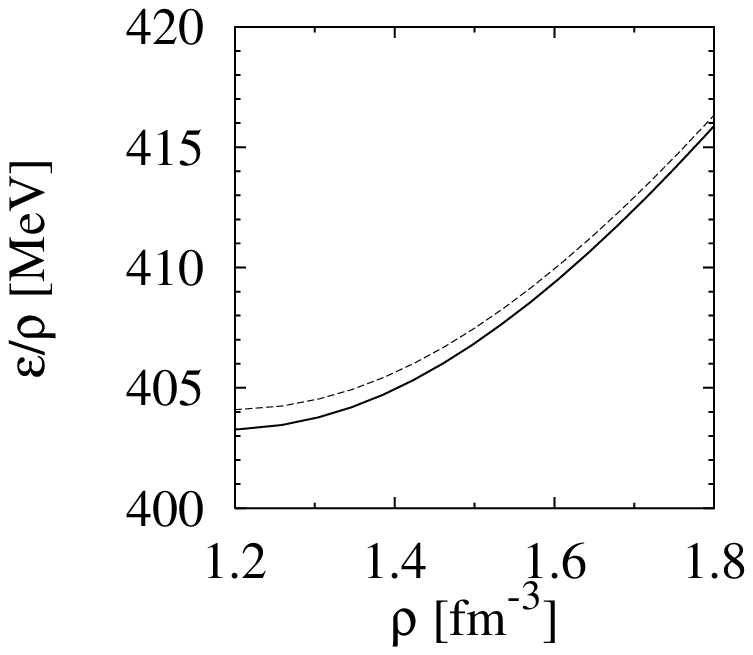,width=7.0cm}
\end{center}
\vspace{-0.5cm}
\caption{\small Left: Number densities of red quarks (solid) and blue quarks
         (dashed) as functions of the quark chemical potential $\mu$.
          Right: Energy per quark as function of the total quark number
          density for a color superconducting system with equal densities
          of gapped and ungapped colors (dashed) and with unequal densities
          as given in the left panel (solid).}
\label{densfig}
\end{figure}

Below the phase transition, the effective chemical potentials
$\mu_r$ and $\mu_b$ are equal to the external chemical potential
$\mu$. At the phase transition point, $\mu_r$ and $\mu_b$ drop by
67~MeV and 51~MeV, respectively and then grow again as functions
of $\mu$. The corresponding  number densities of red and blue
quarks are shown in Fig.~\ref{densfig}. At $\mu=\mu_{crit}$,
$\rho_r$ jumps from zero to $0.42$~fm$^{-3}$, about 2.5 times
nuclear matter density, whereas the density of blue quarks reaches
only twice nuclear matter density at this point, $\rho_b =
0.34$~fm$^{-3}$. Both densities grow of course with increasing
chemical potential, but their difference remains nearly constant.

The unequal densities of red and blue quarks in this state,
although anticipated, clearly deserve further discussion. First of
all we should point out that $\rho_r-\rho_b\neq 0$ is not a result
of our self-consistent treatment, but is already found if we only
take into account the effect of a non-vanishing $\Delta_1$.
Nevertheless, it is usually not discussed in the literature.

Since our model Lagrangian ${\cal L}_{hge}$ is symmetric under
global color $SU(3)$ transformations, the total number of quarks
of each color is a conserved quantity. As in BCS theory, quark
number conservation is violated by the diquark condensates, but
the average numbers are still conserved and given by the densities
$\rho_i$ integrated over the volume. One could therefore ask what
happens if we start with a large but finite system with equal
numbers of red, green and blue quarks at low densities and then
compress it until a color anti-triplet diquark condensate is
formed. According to the above results, in this phase there are
not equally many quarks of each color, but the number of those
quarks which do not participate in the condensate (in the above
case the blue ones) is smaller. Obviously, we cannot get a single
phase with these properties if we start from a system with equal
densities. A possible scenario could be that several domains
emerge in which the symmetry is broken into different directions,
such that the total number of red, green and blue quarks remains
unchanged. Still one could worry about strong forces inside these
domains arising from the non-vanishing net color charge of the
domain. However, we should keep in mind that by construction our
quarks are non-interacting quasiparticles, i.e., at least on the
mean field level there are no forces left.

Alternatively, we can construct a superconducting state with equal
densities for the gapped and ungapped quarks.
To that end we have to introduce different {\it external} chemical potentials
for different red and blue quarks, or, equivalently, an additional external
chemical potential $\mu_8^{ext}$.
Then the second equation in \eq{mu08} becomes
\beq
    \mu_8 \;\= \mu_8^{ext} + 2g_v^{(8)}\rho_8~,
\end{equation}
and in \eq{pot} we should replace $\mu_8^2$ by $(\mu_8 - \mu_8^{ext})^2$.
With this additional external parameter we can enforce the densities of all
colors to be equal, even in the superconducting state.
Obviously this is the case if $\mu_8 = \mu_8^{ext}$.

However, at least within our mean field approximation such a state would be
energetically less favored than a state with the same total density, but
$\mu_8^{ext}$~=~0. This is shown in the right panel of Fig.~\ref{densfig}.
The energy density $\varepsilon$ of the system is given by
\beq
    \varepsilon(T=0,\rho,\rho_8)
    \= \Omega(T=0,\mu,\mu_8) -  \Omega(0,0,0)
       \+ \mu\rho \+ \mu_8^{ext}\rho_8 \,.
\end{equation}
The second term on the r.h.s. has been introduced to normalize the
energy density of the (non-trivial) vacuum to zero. In the right
panel of Fig.~\ref{densfig} the energy per quark,
$\varepsilon/\rho$, is displayed as a function of the total quark
number density $\rho$. The solid line is the result for
$\mu_8^{ext}$~=~0, i.e., it corresponds to the unequal red and
blue quark densities as shown in the left panel. The dashed line
corresponds to the result for equal red and blue quark densities.
As one can see in the figure, the energy of this solution is
always higher than the energy of the solution with unequal
densities. This means, according to this result, a large
homogeneous system of equally many red, green and blue quarks is
unstable against decay into several domains in which the density
of the gapped quarks is larger than the density of the remaining
ungapped quark. On the other hand, the energy difference is not
very large (less than 1~MeV per quark). Therefore it cannot be
excluded that the homogeneous solution with equal densities would
be favored, once effects beyond the mean-field approximation are
taken into account.

\subsection{Lorentz non-invariant interaction}
\label{resultsnli}

In the previous section we found rather small values for the gap parameter
$\Delta_2$. This seems to justify the common practice to neglect this
condensate completely. In order to find out whether the smallness of
$\Delta_2$ is a general feature, we now want to abandon the unnecessary
restriction to a Lorentz invariant interaction, \eq{lii}, and to allow for
unequal values of $g_E$ and $g_M$. Looking at \eq{geff}, we see that the
coupling constants $g_s^{(0)}$, $g_s^{(8)}$ and $h_1$ depend on the sum
$g_E+3g_M$, whereas $g_v^{(0)}$ and $h_2$ depend on the difference $g_E-3g_M$.
$g_v^{(8)}$ is somewhat exceptional because it is mostly given by the
``direct'' interaction term, which is not present in the other channels.
The ``exchange'' term is again proportional to $g_E-3g_M$, but suppressed by
a factor $1/24$.

In principle there are two ways to obtain larger values of
$\Delta_2$. The more dramatic one is to assume that $g_E$ is
larger than $3g_M$. Then the coupling constants $g_v^{(0)}$ and
$h_2$ are positive, i.e., the interaction becomes attractive in
these channels. Indeed, as we will see in
Sec.~\ref{resultsgapless}, an attractive $h_2$ can yield solutions
with large values of $\Delta_2$. However, these solutions might
not be stable, similar to what we found for the schematic example
in the end of Sec.~\ref{disp}. Unfortunately, a detailed study of
the structure of solutions of the coupled gap equations involving
four attractive channels is very difficult and beyond the scope of
our paper.

The second possibility is to make $h_2$ more repulsive. In order
to have a similar starting point as in the previous section, we
keep the sum $g_E+3g_M$ fixed, but increase the difference
$3g_M-g_E$. To see a relatively large effect we take $g_E =
-1.75/\Lambda^2$ and $g_M = 4.25/\Lambda^2$, i.e.,
$3g_M-g_E=14.5/\Lambda^2$, almost three times as large as in the
previous section. As before, we take $m$~=~5~MeV and a cutoff
$\Lambda$~=~600~MeV.
\begin{figure}
\epsfig{file=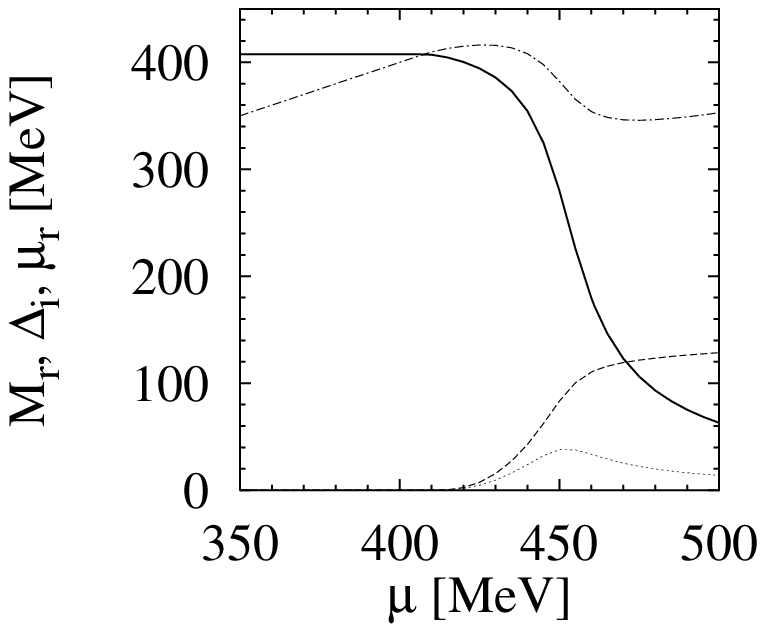,width=7.0cm}
\hfill
\epsfig{file=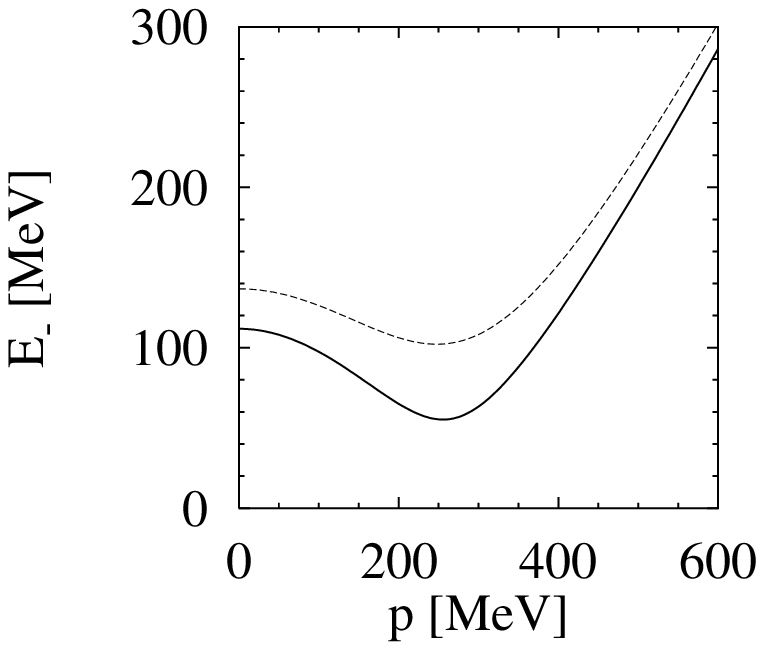,width=7.0cm}
\caption{\small Various quantities obtained with the Lorentz non-invariant
                interaction, $g_E = -1.75/\Lambda^2$ and
                $g_M = 4.25/\Lambda^2$.
                Left panel: $M_r$ (solid), $\Delta_1$ (dashed),
                $-\Delta_2$ dotted, and $\mu_r$ (dashed-dotted)
                as functions of the quark chemical potential $\mu$.
                Right panel: Dispersion law $E_-(\vec p)$ at $\mu$~=~450~MeV.
                The dashed line was calculated neglecting $\Delta_2$ in the
                gap equations, whereas the solid line corresponds to the
                exact solution.}
\label{gapsfigem}
\end{figure}
Our results are displayed in Fig.~\ref{gapsfigem}. In the left
panel we show the behavior of $M_r$, $\Delta_1$, $-\Delta_2$, and
$\mu_r$ as functions of $\mu$. The most striking difference to our
previous example (Fig.~\ref{gapsfig}) is the fact that we now find
a smooth crossover instead of a first-order phase transition.
In the chiral limit the phase transition becomes second order.
A similar effect is known from studies of the chiral phase
transition within the NJL model without diquark condensation.
There it was also found that a first-order phase transition
becomes second order (or a smooth crossover) if the coupling
strength in the (repulsive) vector channel exceeds a certain value
\cite{Bu96}. Although the present model is rather unrealistic
below $\mu \sim$~450~MeV, where it predicts a
color-superconducting quark gas of low density, it is nevertheless
a counter example to the common (and mostly model based) belief
that the chiral phase transition at zero temperature and large
$\mu$ should be first order (see e.g.\@ Refs.~\cite{BeRa99,Halasz99}).

Despite the fact that $|h_2|$ is larger than $h_1$, the absolute
value of $\Delta_2$ remains always smaller than that of
$\Delta_1$. However, unlike in the previous section, $\Delta_2$
can no longer be generally neglected. For instance, at
$\mu$~=~450~MeV, we find $M_r$~=~280.1~MeV, $\mu_r$~=~382.3~MeV,
$\Delta_1$~=~83.2~MeV, and $\Delta_2$~=~-37.8~MeV. In terms of
\eqs{mueff} and (\ref{Deltaeff}) this corresponds to
$\mu_{eff}$~=~384.1~MeV, $M_{eff}$~=~286.9~MeV, and
$|\Delta_{eff}|$~=~55.2~MeV. The resulting dispersion law $E_-(\vec
p)$ is shown in the right panel of Fig.~\ref{gapsfigem} (solid
line). At $|\vec p| = \sqrt{\mu_{eff}^2 - M_{eff}^2}$~=~255.4~MeV it
has a minimum with $E_- = |\Delta_{eff}|$~=~55.2~MeV.
On the other hand, if we neglect $\Delta_2$ in
the gap equation, we get $M_{eff} = M_r$~=~292.5~MeV, $\mu_{eff} =
\mu_r$~=~383.3~MeV, and $\Delta_{eff} = \Delta_1$~=~102.2~MeV.
Consequently, the minimum value of $E_-$ is now 102.2~MeV, almost twice
as much as without neglecting $\Delta_2$.
The corresponding dispersion law is indicated by the dashed line in the
right panel of Fig.~\ref{gapsfigem}. As one can see, the entire
function $E_-(\vec p)$ is shifted to higher energies as compared
with the solid curve and the minimum is more shallow.

\subsection{Gapless color superconductors}
\label{resultsgapless}

Finally, we would like to come back to the question of possible
gapless color superconductors. Obviously, none of our numerical
examples presented so far came close to condition
(\ref{condition}). For instance, if we take $M_r$~=~280.1~MeV,
$\mu_r$~=~382.3~MeV, and $\Delta_1$~=~83.2~MeV, as found in the
previous section at $\mu$~=~450~MeV, one would need $\Delta_2
\simeq$~-113.6~MeV, about three times as large as the actual value. The
situation was even worse in Sec.~\ref{resultsli} where the
discrepancy was about a factor 50 at $\mu = \mu_{crit}$ and became
larger with increasing chemical potential. In fact, none of our
numerical examples fulfilled $|\Delta_2| \geq |\Delta_1|$, which
we identified as a necessary condition for gapless color
superconducting states.

To get some insight, how an interaction could look like which
yields such a state, we can invert the problem and employ the gap
equations to calculate the effective coupling constants which are
consistent with a given set of gap parameters. For instance,
\eq{condition} is obviously fulfilled if we choose $M_r =
\Delta_1$~=~100~MeV and $\mu_r = -\Delta_2$~=~350~MeV. For
simplicity we might assume $M_b = M_r$ and $\mu_b = \mu_r$. Except
of $\Delta_2$ this is within the typical range of these quantities
in the earlier examples. If we now take $m$~=~5~MeV and a cutoff
$\Lambda$~=~600~MeV, as before, and $\mu$~=~450~MeV, the gap
equations yield $g_s^{(0)}\Lambda^2 = 3.36$, $g_v^{(0)}\Lambda^2 =
-1.41$, $h_1\Lambda^2 = 6.80$, $h_2\Lambda^2 = 6.18$, and
$g_s^{(8)}\Lambda^2 = g_v^{(8)}\Lambda^2 = 0$. Here the essential
difference to our earlier examples is the need of an attractive
interaction in the $h_2$ channel. Furthermore, the interaction is
relatively strong in both diquark channels. However, for these
parameters there is another solution with $M_r = M_b$~=~58.1~MeV,
$\mu_r = \mu_b$~=~362.8~MeV, $\Delta_1$~=~966.6~MeV, and
$\Delta_2$~=~16.1~MeV. In order to decide which of the two
solutions is the energetically favored one we have to evaluate the
thermodynamic potential. It turns out that for the gapless
solution the value of $\Omega$ is about 900 MeV/fm$^3$ higher than
for the other solution. Hence, similar to what we found in the
schematic example of Sec.~\ref{disp}, the gapless state does not
correspond to a stable solution. In fact, we did not succeed to
construct a stable color-superconducting solution. This might
indicate that such a state does not exist, although a rigorous
proof is still missing.

\section{Conclusions}
\label{conclusions}

We analyzed the ground state properties of an isotropic two-flavor
color-superconductor at finite quark chemical potential within an
NJL-type model. In general, a self-consistent treatment of this
problem requires to consider several condensates which are usually
neglected: The breakdown of Lorentz invariance in dense systems
implies the possible existence of Lorentz non-invariant diquark
condensates, while the spontaneous breaking of color $SU(3)$ in a
color superconductor naturally leads to the existence of $SU(3)$
non-invariant quark-antiquark condensates. We found that at least six
different expectation values (two diquark condensates and four
quark-antiquark condensates, \eqs{delta1} to (\ref{rho8})) have
to be taken into account in a self-consistent calculation and we
derived a set of six coupled gap equations for these expectation
values.

The actual importance of the various condensates depends of course
on the interaction. Since at present not very much is known about
the effective interaction which describes the deconfined phase at
moderate densities, this leaves room for surprises and possible
new phenomena. For instance, in the numerical example discussed in
Sec.~\ref{resultsli} the chiral phase transition at finite $\mu$
and zero temperature was of first order, in agreement with the
general expectation. However, in Sec.~\ref{resultsnli}, where a
Lagrangian with strong repulsive interactions was chosen, we found
a smooth crossover. Similarly, the effect of the Lorentz
non-invariant diquark condensate $\delta_2$, which is usually
neglected, was indeed found to be small in Sec.~\ref{resultsli},
but relatively large in Sec.~\ref{resultsnli}. There it caused a
reduction of the effective gap parameter $|\Delta_{eff}|$, (the
minimum energy of a quasiquark excitation) by almost 50\%. In
general, the numerical values of the condensates depend also on
how the divergent integrals are handled. In this article we used a
sharp cutoff. A comparison with other approaches is yet to be
done.

We also discussed the possibility of ``gapless color
superconductors'', i.e., states with non-vanishing diquark
condensates, but $|\Delta_{eff}|$~=~0. We have seen that such
states can in principle exist, provided the gaps of the various
condensates are related to each other in a certain way (see
\eq{condition}). In practice, however, all gapless superconducting
states we constructed turned out to be unstable solutions of the
gap equations. A similar observation was made in
Ref.~\cite{ABeR00} for gapless states in the color-flavor locked
phase. This suggests that gapless color superconductors might be
in general unstable.

As a consequence of the spontaneous color $SU(3)$ breaking through the
diquark condensates, there are also $SU(3)$ non-invariant quark-antiquark
condensates in a color superconductor. For instance, the familiar scalar
condensate $\qq$ is in general different for quarks which participate in
the diquark condensate (``red'' and ``green'') and those which do not
(``blue''). This also leads to different constituent quark masses for
gapped and ungapped quarks. However, at least in our numerical examples
this difference turned out to be quite small.
On the other hand we found that (for equal chemical potentials)
the density of the gapped quarks was considerably larger than the density
of the ungapped quarks. Since the total number of red, green and blue quarks
should be equal in a finite system, this could lead to the emergence of
domains in which color $SU(3)$ is broken into different directions.
\\
\hspace{5mm}

{\bf\large Acknowledgments}

Part of this work was done during the collaboration meeting on color
superconductivity in Trento. We thank ECT$^*$ for financial support
during this meeting and the organizer, Georges Ripka, and all
participants for stimulating discussions.  Two of us (M.B. and J.H.)
also acknowledge financial support by INT in Seattle during the
program ``QCD at Finite Baryon Density'' at a very early stage of this
work. One of us (M.O.) acknowledges support from the Alexander von
Humboldt-foundation as a Feodor-Lynen fellow. This work was in part
supported by the BMBF.


\end{document}